\documentstyle[11pt,newpasp,twoside,epsf]{article}
\def\kms{km\,s$^{-1}$}

\def\feii{Fe\,{\sc ii}}
\def\caii{Ca\,{\sc ii}}
\def\mgii{Mg\,{\sc ii}}
\def\mgi{Mg\,{\sc i}}
\def\hi{H\,{\sc i}}

\def\civ{C\,{\sc iv}}

\def\edcomment#1{\iffalse\marginpar{\raggedright\sl#1\/}\else\relax\fi}
\reversemarginpar

\begin{document}
\title{An~MHD-driven~Disk~Wind~Outflow~in~SDSS~J0300+0048?}
\author{Patrick B. Hall}
\affil{Princeton University Observatory, Princeton, NJ 08544-1001, USA, and
Departamento de Astronom\'{\i}a y Astrof\'{\i}sica, Facultad de F\'{\i}sica,
Pontificia Universidad Cat\'{o}lica de Chile, Casilla 306, Santiago 22, Chile}
\author{Damien Hutsem\'ekers}
\affil{Research Associate FNRS, University of Li\`ege,
     All\'ee du 6 ao\^ut 17, Bat. 5c, 4000 Li\`ege, Belgium}

\begin{abstract}
The outflow in SDSS J0300+0048 has the highest column density yet reported
for a broad absorption line quasar. 
The absorption from different ions is also segregated as a function of velocity
in a way that can only be explained by a disk wind outflow.  Furthermore,
the presence of the such large column densities of gas at the high observed
outflow velocities may be incompatible with purely radiative acceleration.
MHD contributions to the acceleration should be considered seriously.
\end{abstract}

\section{Observations and Implications}

\begin{figure}
\plotfiddle{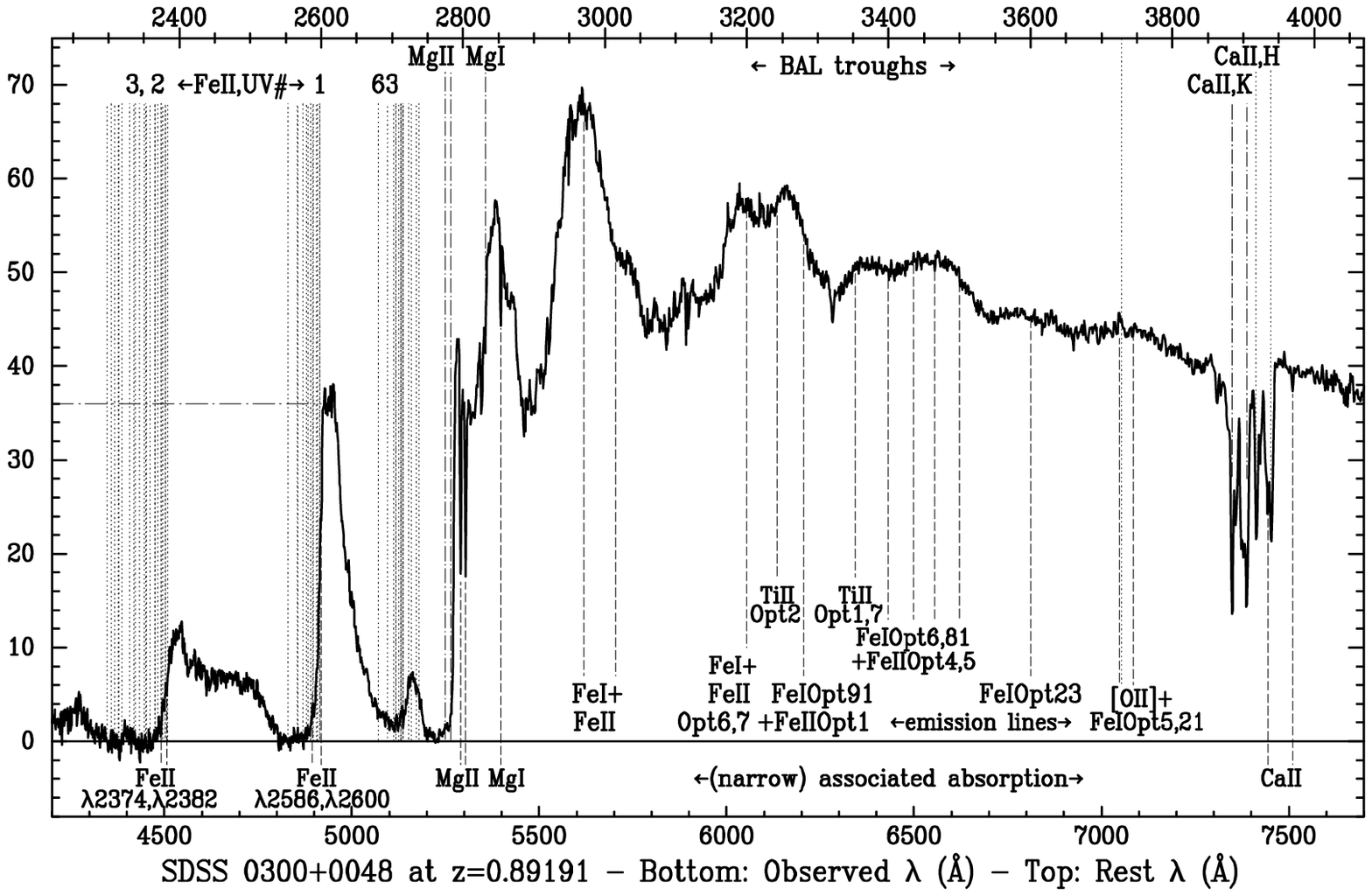}{3.35in}{0.0}{82.5}{82.5}{-265}{-205}
\caption{Low-resolution spectrum of SDSS J0300+0048.}
\end{figure}

\begin{figure}
\plotfiddle{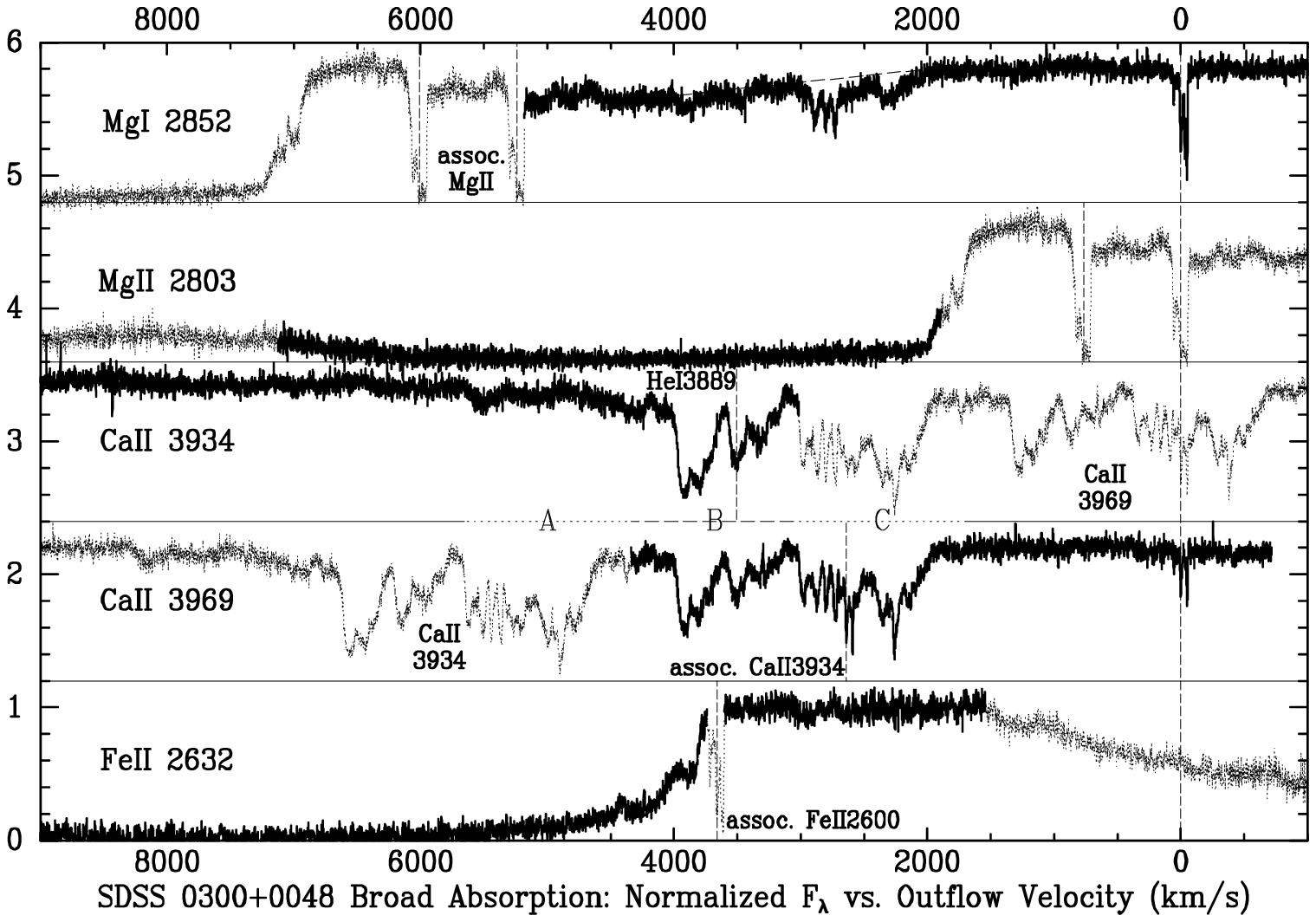}{3.5in}{0.0}{87.}{87.}{-275}{-220}
\caption{Portions\,of\,the\,high-resolution\,spectrum\,of~SDSS\,J0300+0048.}
\end{figure}

SDSS J030000.56+004828.0 
(Fig.~1) is a luminous, ``overlapping-trough'' broad absorption
line (BAL) quasar (Hall et al.\ 2002) and one of only a handful of \caii\ BAL
quasars known.  High-resolution spectroscopy has been obtained for this object
using the Ultraviolet Visual Echelle Spectrograph
on the ESO VLT (Hall et al.\ 2003).
Figure 2 shows the UVES spectrum around five important transitions,
with outflow velocity increasing to the left.
The \caii, \mgii, and \mgi\ column densities in this object are the largest
reported to date for any BAL outflow.  The column density of metals is $\sim$200
times higher than in the only other \caii\ BAL studied at high resolution,
QSO J2359$-$1241 (Arav et al.\ 2001).

Figure 3 sketches the inferred structure of the outflow in SDSS J0300+0048.
The large column density of \caii\ observed in this object can only exist in
gas well shielded by an H\,{\sc i} ionization front (Ferland \& Persson 1989),
and \caii\ is only seen at the lowest line-of-sight velocities in the outflow.
Therefore, the lowest-velocity outflowing gas is farthest from the quasar.
This result is very unlikely if BAL outflows are nearly spherical ``cocoons''
of dusty gas.  On the other hand, it is easily understood in disk-wind models
where the gas originating closest to the quasar is accelerated to the highest
velocities, and where we can look across the streamlines of the outflow instead
of just down along them.  
There is considerable other evidence for a disk wind outflow in this object:

\noindent{$\bullet$~looking across the streamlines can explain why the lowest
velocity gas in the outflow is ``detached'' (blueshifted from the systemic
redshift) by $\sim$1650 km~s$^{-1}$ --- the gas undergoes acceleration
before it crosses our line of sight.

\noindent{$\bullet$~The lowest velocity BAL region produces strong
\caii\ absorption but 
no significant excited \feii\ (\feii*) absorption, while the higher
velocity excited \feii\ absorption region produces very little \caii.
Figure 3 shows how this segregation can arise naturally in a disk wind model.
The density at the base of the wind (the surface of the accretion disk)
increases with radius over the relevant range of radii, at least in a standard
thin disk model (Shakura \& Sunyaev 1973).
Thus, both the \feii* and \caii\ BAL regions likely have densities high enough
to populate excited levels of \feii, but the \caii\ BAL region must have a
temperature low enough ($T$$<$1000\,K) to prevent them from being
significantly populated.
Outflowing clouds in a ``cocoon'' cannot explain this segregation
unless the lowest velocity clouds are farthest from the quasar (which would
be rather contrived for a cocoon model).  Otherwise, gas shielded by optically
thick, low-velocity \caii\ clouds would produce \caii\ absorption at high
velocities as well.

\begin{figure}
\plotone{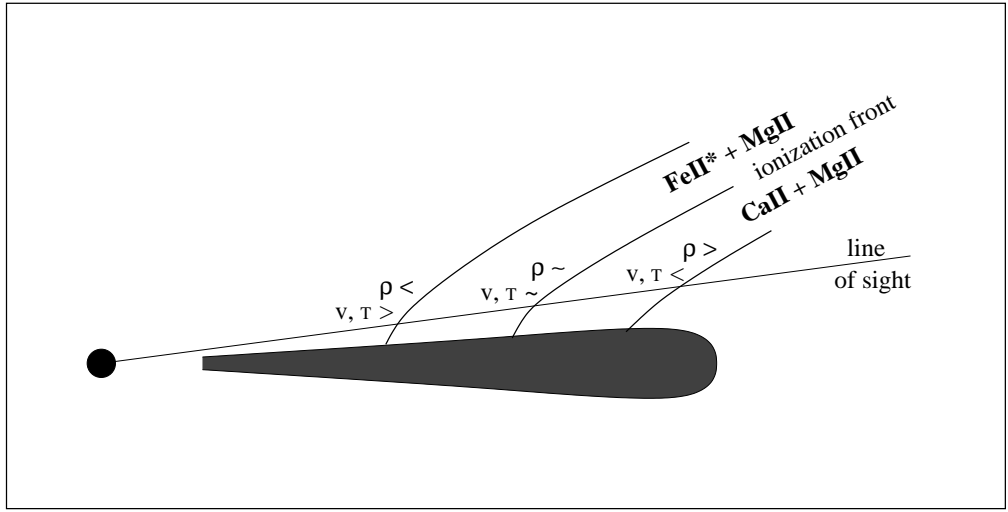}
\caption{The thin line shows our sightline to the black hole and central
continuum source (black dot).  The curved lines show the streamlines
of wind from the accretion disk (half of which is shown, in profile,
as the elongated slab).  The symbols beside each streamline show how the
density ($\rho$), velocity ($v$) and temperature ($T$)
increase ($>$), decrease ($<$), or remain the same ($\sim$) 
{\em between} streamlines along the line of sight
(changes different from those {\em along} an individual streamline,
where $v$ and $T$ 
always increase and $\rho$ 
always decreases with increasing radial distance).
In this Figure, the \hi\ ionization front is assumed to be coincident with a
streamline, 
though that may not be the case in reality.
Inside this front around the quasar we observe \mgii\ and \feii* in absorption,
while outside it we observe \mgii\ and \caii\ in absorption.}
\end{figure}

This model predicts that the \mgii\ absorption at velocities 
$v$$<$4000\,\kms\ should not be accompanied by \civ\ absorption.  Such 
segregation of \mgii\ and \civ\ is not usually observed in quasars.
However, that could be because most lines of sight through BAL quasar outflows 
are less highly inclined relative to the outflow streamlines than is our line
of sight in SDSS J0300+0048.  Since acceleration occurs along the streamlines,
lines of sight oriented along them
may tend show all ions present in absorption at a range of velocities.

Finally, the acceleration of a large column of velocity-segregated, very
low-ionization gas to the observed velocities of up to 4000\,\kms\ may require
more than continuum and line radiation pressure on atoms and ions.
Radiation pressure on ionized gas can push accompanying neutral gas along in an
outflow, but that scenario would not produce the velocity-segregated outflow
seen in SDSS J0300+0048.
Continuum radiation pressure on neutral atoms is unlikely to produce 
an acceleration comparable to that from line pressure on ionized gas,
but radiation pressure on dust in the low-ionization zones of the outflow
may produce significant acceleration (Dopita et al.\ 2002).

Magnetohydrodynamical effects (e.g., magnetocentrifugal acceleration)
can also contribute to the acceleration of gas
even if the ionization fraction is very small.
Moreover, we note that MHD acceleration would be {\em required} if the
X-ray absorbing column of $N_H\simeq3.5\times10^{24}$~cm$^{-2}$ in this object
(Hall et al., in preparation) is outflowing rather than stationary.
Even assuming 100\% covering and absorption of 100\% of the radiation
of a quasar radiating at the Eddington limit, a column density of gas
$N_H>1.43\times10^{24}$~cm$^{-2}$ cannot be accelerated away from a quasar
by radiation alone (Hamann et al.\ 2002).

Unambiguous evidence of MHD driving requires finding
a BAL quasar with X-ray absorbing column $N_H>1.43\times10^{24}$~cm$^{-2}$
{\em and} narrow, detached absorption troughs
which trace the X-ray absorbing column.  Detecting that high a column
in outflow (rather than at the systemic redshift) would unambiguously point to
MHD acceleration in quasar outflows.  SBS 1542+541 shows detached UV troughs of
ions up to Si\,{\sc xii} (Telfer et al.\ 1998), but its
$N_H \leq 10^{23}$~cm$^{-2}$ does not {\em require} MHD acceleration.
More promising may be the recently discovered detached X-ray BAL troughs 
(from ions up to Fe\,{\sc xxvi}). 
The highest column density reported among 4 or 5 such objects is 
$N_H \simeq 5.7 \times 10^{23}$~cm$^{-2}$ (PDS~456; Reeves et al.\ 2003),
within a factor of 2.5 of that needed to confirm MHD driving.

\acknowledgements
Funding for the creation and distribution of the SDSS Archive
(http://www.sdss.org/) has been provided by the Alfred P. Sloan
Foundation, the Participating Institutions, the National Aeronautics
and Space Administration, the National Science Foundation, the
U.S. Department of Energy, the Japanese Monbukagakusho, and the Max
Planck Society.

\end{document}